\documentclass[10pt]{article}
\usepackage[letterpaper]{geometry}
\usepackage{hicss}
\usepackage{times}
\usepackage{hyphenat}
\usepackage{url}
\usepackage{latexsym}
\usepackage{indentfirst}
\usepackage{graphicx}
\usepackage{hyperref}
\graphicspath{{images/}}
\usepackage{makecell}
\usepackage{adjustbox}
\usepackage{wrapfig}
\usepackage{float}
\usepackage{pdfpages}
\usepackage{enumitem}
\hypersetup{
    colorlinks,
    linkcolor={red!50!black},
    citecolor={blue!50!black},
    urlcolor={blue!80!black}
}

\usepackage{paralist}
\usepackage{booktabs}

\newcommand{\mypar}[1]{\smallskip\vspace{-0.1em}\noindent\textbf{#1.}}

\newcommand\mytitle{What is Business Process Automation Anyway?}

\title{\mytitle}

\author{Hoang Vu \\
  SAP \\
  {\underline{h.vu@sap.com}} \\\And
  Henrik Leopold \\
  Kühne Logistics University \\
  {\underline{ henrik.leopold@the-klu.org} }\\\And 
  Han van der Aa \\
  University of Mannheim \\
  {\underline{han.van.der.aa@uni-mannheim.de}} \\}

\date{}

\begin{document}
\maketitle
\begin{abstract}
Many organizations strive to increase the level of automation in their business processes. While automation historically was mainly concerned with automating physical labor, current automation efforts mostly focus on automation in a digital manner, thus targeting work that is related to the interaction between humans and computers. This type of automation, commonly referred to as business process automation, has many facets. Yet, academic literature mainly focuses on Robotic Process Automation, a specific automation capability.
Recognizing that leading vendors offer automation capabilities going way beyond that, we use this paper to develop a detailed understanding of business process automation in industry. To this end, we conduct a structured market analysis of the 18 predominant vendors of business process automation solutions as identified by Gartner. As a result, we provide a comprehensive overview of the business process automation capabilities currently offered by industrial vendors. We show which types and facets of automation exist and which aspects represent promising directions for the future.
\end{abstract}
\vspace*{-5mm}
\noindent \textbf{Keywords:} Business process automation, RPA, Workflow management, Market analysis

\section{Introduction}
\label{sec:introduction}

The desire of organizations to automate human labor has a long history. In the context of factories and, more specifically, manufacturing, typically the invention and development  of the steam engine in the $18^{th}$ century is considered the major turning point \cite{atack2008steam}. After the introduction of electricity, the use of punch cards allowed to control and automate machines and perform work not only faster but also more precisely \cite{acemoglu2018artificial}. While  automation in the context of manufacturing is an ongoing topic with numerous innovations \cite{lu2020smart}, the scope of automation nowadays is much broader. In particular, current work automation efforts commonly focus on automation in a digital manner, targeting work  
 that is related to the interaction between humans and computers \cite{parasuraman2000model}. 

In this paper, we particularly focus on such digital initiatives for work automation, separating it from automation through physical means, such as automated assembly lines. 
In the remainder, following industry convention, we shall refer to such digital automation of work as \textit{business process automation}.
While early technology in this regard was mainly concerned with the management and coordination of work in the context of workflow systems \cite{van2004workflow}, more recent technologies, such as Robotic Process Automation (RPA), aim to fully automate certain tasks humans perform using so-called bots \cite{ivanvcic2019robotic}.
When considering academic literature, it is notable that RPA dominates the current discourse on business process automation. In fact, recent literature in this area is almost exclusively concerned with RPA~\cite{ivanvcic2019robotic,syed2020robotic,hofmann2020robotic,enriquez2020robotic}. While this is not a problem per se, it is important to recognize that business process automation is by no means limited to RPA. This becomes evident when looking  at the product portfolio of leading RPA vendors, such as UiPath, Automation Anywhere, or Microsoft, where RPA is just one of many automation capabilities they offer. This diversity, however, is currently not reflected in the academic discourse. Furthermore, there is little understanding of the different types and facets of business process automation.  

Recognizing this, we use this paper to develop a comprehensive understanding of business process automation in industry. More specifically, we set out to answer the following three research questions:

\begin{compactenum}[RQ1:]
 \item What are the main process automation types and associated capabilities offered in industry?
 \item How do vendors support organizations in operationalizing their automation efforts?
 \item How well are the various automation capabilities covered by industry vendors?
\end{compactenum}

\noindent To answer these questions, we conducted a structured market analysis of the 18 predominant vendors of business process automation solutions as identified by Gartner. As a result, we provide a comprehensive overview of the business process automation capabilities currently offered by industrial vendors. 

The rest of the paper is structured as follows. \autoref{sec:background} provides a brief introduction into the topic of automation and an overview of related work. \autoref{sec:methodology} explains our research methodology. \autoref{sec:automation_types} and \autoref{sec:automation_support} present the results on automation types and automation support, respectively. \autoref{sec:vendor_coverage} reflects on the different vendor profiles and their coverage before \autoref{sec:conclusion} concludes the paper.

\section{Background}
\label{sec:background}

The term \textit{automation} is used in various ways and in a wide range of contexts in both academic literature and spoken language~\cite{save2012designing,wickens2018automation,parasuraman2000model}. In the Oxford English Dictionary (2006), automation is defined as:
\begin{compactenum}
    \item Automatic control of the manufacture of a product through a number of successive stages;
    \item the application of automatic control to any branch of industry or science;
    \item by extension, the use of electronic or mechanical devices to replace human labor.
\end{compactenum}

\noindent This definition emphasizes that automation is largely related to \textit{automatic control} and can refer to electronic as well as mechanical action~\cite{parasuraman2000model}. In this paper, we are concerned with human work that is automated by means of software in the context of business processes. We therefore build on the definition from Parasuraman and Riley~\cite{parasuraman1997humans} that emphasizes the interaction between humans and computers and define automation as ``\textit{the partial or full execution of a task by a computer that has been previously carried out by a human}''.  

Literature on such automation can be divided into three main categories: workflow management, robotic process automation, and robotic process mining. 


\textit{Workflow Management} is concerned with providing automated support for the execution of business processes~\cite{ouyang2015workflow}. Research in this area has addressed various aspects including workflow design patterns~\cite{van2003workflow}, conceptual approaches~\cite{medina1993action}, workflow systems~\cite{reichert2003adept}, and workflow modeling~\cite{van2005yawl}. Most literature on workflow management stems from the 1990s and early 2000s, indicating that academic discourse has moved on.    

\textit{Robotic process automation (RPA)} relates to software-based technology that aims to fully automate certain tasks humans perform at a computer \cite{ivanvcic2019robotic}. The core idea behind RPA is to mimic the user's mouse and keyboard interactions and, in this way, automate tasks that are repetitive and follow regular execution patterns. Literature on RPA has mainly analyzed and discussed the importance and potential of RPA in industry~\cite{aguirre2017automation,van2018robotic,hofmann2020robotic}. Other authors have also focused on the organizational challenges coming with the introduction and use of RPA~\cite{willcocks2015function,syed2020robotic} and the capabilities of RPA are actually used by organizations~\cite{enriquez2020robotic}. Recognizing that RPA needs to be integrated into existing automation efforts, there is also an ongoing discussion on how RPA can be combined with existing efforts in the area of business process management \cite{leshob2018towards,mendling2018machine,konig2020integrating}. 
    
\textit{Robotic process mining (RPM)} is a very recent research domain~\cite{leno2019action}. The goal of RPM techniques is to automatically identify automatable routines based on user interaction (UI) logs. These logs capture the keyboard and mouse interactions of the user including additional information, such as the caption of buttons or the labels of input fields the user interacted with. RPM can therefore be considered as an enabler for RPA since it supports organizations in identifying \textit{what} can be automated. Research belonging to this area typically focuses on defining specific techniques that identify automatable routines in UI logs~\cite{leno2020identifying,agostinelli2021exploring} or facilitate the creation of UI logs~\cite{leno2019action}.

This brief review of literature from the area of business process automation shows that the current discourse on automation almost exclusively focuses on RPA and RPA-related techniques. Given that leading vendors offer much more than RPA-related solutions, we are convinced it is both useful and necessary to develop a more comprehensive understanding of the facets of automation. In the next section, we explain how we will approach this from a methodological perspective.

\section{Methodology}
\label{sec:methodology}

To answer the three research questions posed above, we conducted a structured market analysis. 
Based on the Gartner Magic Quadrant for Robotic Process Automation\footnote{\url{https://www.gartner.com/en/documents/4004033}}, we identified 18 predominant vendors for RPA and process automation solutions in the market. The rationale of our analysis was that an analysis of the capabilities offered by these vendors provides a comprehensive picture of the different facets of business process automation in industry. 

To obtain the desired insights from the considered 18 vendors, we analyzed and coded their websites using a specific methodology for coding qualitative data~\cite{Saldana2009}. 
Our coding procedure consisted of three main steps: 1) hypothesis coding, 2) holistic coding, and 3) category derivation. During the \textit{hypothesis coding}, we created a set of codes we expected to find in the data without actually conducting any further analysis. Among others, this included codes such as ``\textit{automation type}'' or ``\textit{automation technology}''. During the \textit{holistic coding}, we identified possible categories that could emerge from and represent the data. We realized that one of the key challenges was to understand which range of terms vendors would use to refer to similar or even identical capabilities. For instance, so-called ``\textit{task recording}'', which supports the development of RPA bots, is also referred to as ``\textit{process recorder}'', ``\textit{discovery bot}'', or ``\textit{screen recorder}''. Finally, during the \textit{category derivation}, we compared both coding schemes and analyzed how they relate to each other in order to achieve a more concise representation of the identified categories. As a result, we refined but also merged a number of codes to arrive at the final categorization, leading to 21 capabilities,  each offered by at least 2 vendors. 

In the following sections, we use these capabilities and their respective coverage by vendors in order to answer the three research questions raised in \autoref{sec:introduction}.


\section{Automation Types}
\label{sec:automation_types}
In this section, we answer RQ1: \textit{what are the main process automation types and associated capabilities offered in industry?}
Through the market analysis described in \autoref{sec:methodology}, we identified four high-level types of process automation being offered: 
(1) \textit{Robotic Process Automation (RPA)}, which automates individual process steps on top of existing technical infrastructure, (2) \textit{workflow automation}, which automates the orchestration of a business process, (3) \textit{Integration Platform as a Service (iPaas)}, which automates interactions between different systems that support a process, and (4) \textit{integrated process automation}, which provides solutions that encompass the prior automation types where applicable, providing integral automation solutions for business processes.

\begin{figure}[h]
    \centering
    \includegraphics[width=\linewidth]{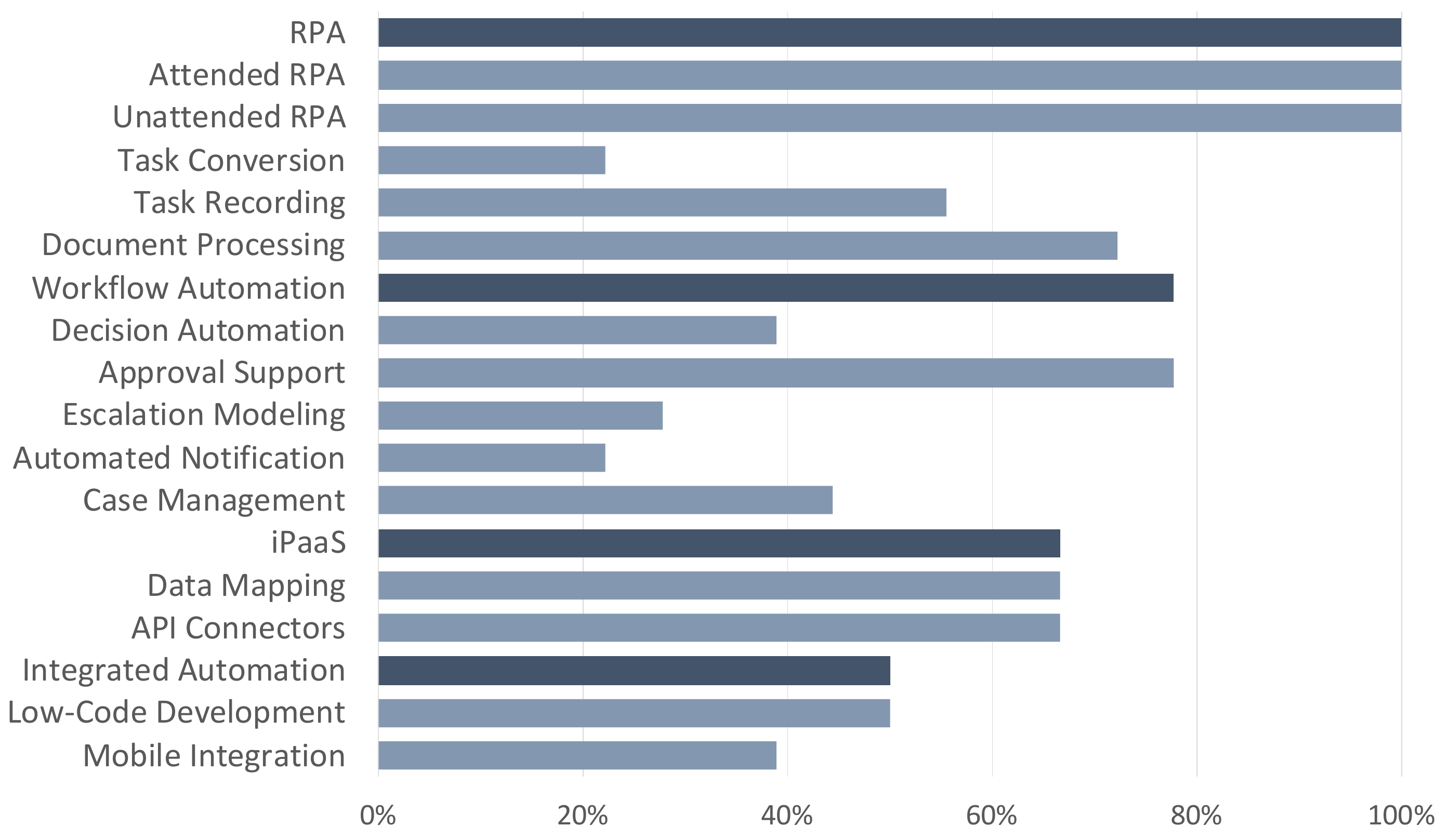}
    \caption{Vendor coverage of automation types and associated capabilities}
    \label{fig:types}
\end{figure}

\autoref{fig:types} visualizes the indicated coverage of the 18 vendors included in our study with respect to these automation types, as well as the key capabilities underlying them. We discuss the specifics of these types and capabilities in the remainder of this section.


\subsection{Robotic Process Automation (RPA)}
\label{sec:types:rpa}

Many business processes include repetitive tasks that have to be carried out by the employees of a company on a daily basis. To free up the time of the human worker to focus on knowledge-based work within their company, RPA was introduced to emulate the human actions on a software or digital systems and with that replace the human worker in performing repetitive tasks. 

In terms of capabilities, we observed that RPA vendors primarily characterize their offering based on two general types of RPA bots (\textit{attended} and \textit{unattended}), as well as additional capabilities in terms of development support through \textit{task conversion} and \textit{recording}, and \textit{document processing} technology to broaden the functionality of bots.

\mypar{Attended \& Unattended RPA bots} 
RPA bots can, at a high level, be characterized as being either attended or unattended.
An \textit{attended RPA bot} needs to be explicitly invoked by a user when it needs to perform a task, such as a user that indicates that the contents of an invoice should be extracted and put into the appropriate system.
Attended RPA bots typically run on the employees' computer or a dedicated server. 
In contrast, an \textit{unattended RPA bot} does not require user interaction and, rather, runs back-office processes. Therefore, unattended bots are usually deployed on workstations or servers, so that they  run independently from the availability of employees. Unattended RPA bots either operate on a scheduled basis, where during their design, a time interval (e.g., every day at 9:00am) is set for the bot to run the back-office processes (such as sending out recurring managerial reports), or on a triggered basis, meaning that a preceding process step triggers the bot via an API call (e.g., once an invoice has been paid, an RPA bot automatically archives it). 

Given the widespread relevance of both attended and unattended RPA, as well as the selection criteria employed by Gartner in its market analysis, all considered vendors offer both types of RPA bots.

\mypar{Task Conversion} 
Task converters can be offered by vendors to support users in the development of RPA bots. These converters
aim to extract  information about an automation flow, captured in some form of documentation, and  turning it into an RPA bot automatically. An example could be a textual or (semi-)formal document that describes the steps that need to be taken to assess the completeness of a received purchase order, and turning this information into a bot that automates this assessment task. According to our vendor analysis, only around 20 percent of vendors offer this capability. 

\mypar{Task Recording} 
An additional, more common capability to support the development of RPA bots is task recording. This capability allows the recording of a users' computer screen and captures every interaction a user performs with their computer in detail, for example, inserting data, clicking a button or pressing enter. This sequence of recorded steps is then  automatically translated into an RPA bot that can perform them on behalf of users. 
A particular strength here is that task recorders can capture interactions with any application that a user accesses during the recorded process, including websites and local applications. Based on our analysis, about 55 percent of vendors provide this capability to their users.

\mypar{Document Processing}
Finally, to broaden the applicability of RPA bots, vendors offer document processing technology as a capability, which allows RPA bots to extract unstructured data from documents, such as scans of received invoices, so that it can be used in subsequent steps in a business processes. Such document processing is typically performed using Optical Character Recognition (OCR). The extracted data then gets structured and labeled so that an RPA bot can leverage it. About 70 percent of automation vendors have this capability in their automation portfolio.

\subsection{Workflow Automation}
\label{sec:types:workflowautomation}

Whereas RPA is concerned with the automation of specific steps within a process, \textit{workflow automation} instead aims to automate the overall coordination of a process, achieved by, e.g., automatically assigning work items to employees and the routing of process instances according to business rules.
In this regard, key capabilities offered by vendors are: decision automation, approval support, escalation modeling, automated notifications, and case management.


\mypar{Decision Automation}
Some automation vendors offer \textit{decision automation} to enable the modeling of business rules within a company. By setting up these artifacts through a decision table that validates for example certain thresholds or determines the appropriate user group for a specific task, this capability allows the incorporation of business logic into their processes. Based on the analysis, around 40 percent of vendors provide this capability to their customers. 

\mypar{Approval Support}
Given that not all decisions can be automated by business rules, vendors also offer \textit{approval support} as a capability that allows managers to efficiently handle and respond to approval requests by accepting, rejecting, or sending them back. Based on the managers decision of an approval the workflow can be modeled into separate paths. This capability is covered by around 80 percent of automation vendors in the market.


\mypar{Escalation Modeling}
Even when workflow orchestration is automated, certain tasks may still not be performed in a timely manner, such as approval requests not being processed, or not being executed at all. For these types of scenarios an escalation workflow can be modeled to handle these cases and for example escalate to a higher management level or inform the central IT department about the process instance. This capability ensures the handling of errors and a timely execution of processes. Based on the vendor analysis, around 30 percent of automation companies offer this capability to their users.

\mypar{Automated Notification}
When a workflow requires the interaction of employees or managers, they usually have to be informed whenever a user task needs to be executed by them. This notification allows them to pick up a user task without having to actively monitor the process themselves. Simply by receiving a notification on their computer or mobile phone all required information for the process are shown to them and they can perform the user task from their device. According to the analysis, around 20 percent of vendors provide a unified notification capability in their automation offering.

\mypar{Case Management}
Many processes require knowledge workers to gather data and make decisions on it in a timely and collaborative way. An example where case management is utilized is the emergency reception process in hospitals. Each accident that gets delivered into a hospital can be different to another and with that each case needs to be examined individually by an expert and therefore each action is derived from an expert opinion. Most of the times, these cases require collaboration between multiple doctors from different fields and have to be handled in a timely manner. For this reason the \textit{case management} capability can support these case workers by providing the right data in a collaboration platform to quickly make a joint decision for action. Based on our market analysis, around 45 percent of the vendors offer this capability in their product portfolio. 

\subsection{Integration Platform as a Service (iPaaS)}

The execution of business processes often spans multiple information systems, such as an order-to-cash process that relates to both order-handling and payment systems.
To ensure a smooth and faultless process flow in such settings, these systems have to be connected to each other. In this context, iPaaS is often used to connect two systems indirectly with each other via API, thus automating interactions between systems. The two main automation capabilities that vendors provide in this regard are data mapping and API connectors.

\mypar{Data Mapping}
A sending system forwards data towards this platform which then transforms and routes the data towards a target system automatically. Therefore, it usually comprises multiple data mapping techniques to handle complex data structures in multiple formats, such as JSON or XML. Transformation of the data by validating each line item or enriching each segment based on business logic are some examples of the \textit{data mapping} capability. In most cases the sending and receiving systems require different data structures, which is why this capability is intensively used for each integration scenario. Around 70 percent of vendors offer this capability to their customers.

\mypar{API Connectors}
Similar to the data mapping requirements usually the sending and receiving systems use different technical protocols when sending and receiving data. This requires different connector technologies based on the respective systems involved such as REST, OData or SOAP. An example would be converting a flat text file received in an SFTP server into a format and enriching it so that a REST interface can handle the data. The integration platform as a service usually includes an API management technology to manage the security behind each API call and also a connector framework to  develop additional connectors to different systems and technologies. According to our analysis, about 70 percent of automation vendors offer this capability to the market.

\subsection{Integrated Process Automation} 

Whereas the preceding types all focused on the automation of individual aspects of a business process, \textit{integrated process automation} targets the automation of a business process in a manner that encompasses the former parts, i.e., the automation of individual tasks, their orchestration, and technical integration. The aim of this type of automation is to automate processes to the extent that this is possible, commonly supported by low-code development tools that combine RPA, AI, business rules, and human interaction. This makes low-code development a crucial capability in this context, whereas various vendors also specifically offer support for mobile integration.


\mypar{Low-Code Development} 
The capability \textit{low-code development} focuses mainly on application development for computer and web-browser applications. Instead of providing the developer only with a development environment, it provides pre-built coding blocks that a user can drag and drop when building their user interface. This lowers the entry level in terms of technical expertise required to start building a user interface and therefore targets a broader range of users compared to traditional development environments. In that regard, this capability is crucial to make integrated process automation accessible to organizations.
According to our analysis, all vendors that offer integrated process automation solutions, around 50 percent of the total, incorporate this capability.  

\mypar{Mobile Integration} 
As an addition to low-code development,  this capability focuses on application development tailored to smartphones. 
Although the development experience can be similar to low-code application development, the use cases that are targeted 
for mobile users can differ, due to the smaller interface and usage mode of these devices. 
Therefore, those parts of a process handled by smartphones are usually focused on providing only the key information and functionality to users in a convenient manner. An example for a mobile application is a management approval user interface where a manager only sees the most relevant information and can simply approve or reject an approval task from their mobile device. Based on the market analysis, about 40 percent of vendors offer this capability to their customers.

\section{Automation Support}
\label{sec:automation_support}
In this section we focus on RQ2: \textit{How do vendors support organizations in operationalizing their automation efforts}? 
For this, we focus on the capabilities that automation vendors provide to their users that go beyond the actual task of automating (parts of) a business process. First, we cover \textit{opportunity analysis}, which targets the early stages of an automation project that revolve around the identification of automation opportunities and the requirements associated with the implementation of these opportunities. Second, we consider \textit{automation management}, which focuses on rolling out automation implementations within an organization, e.g., through proper governance, and their subsequent life-cycle management, e.g., through continuous monitoring.

\begin{figure}[htbp]
    \centering
    \includegraphics[width=\linewidth]{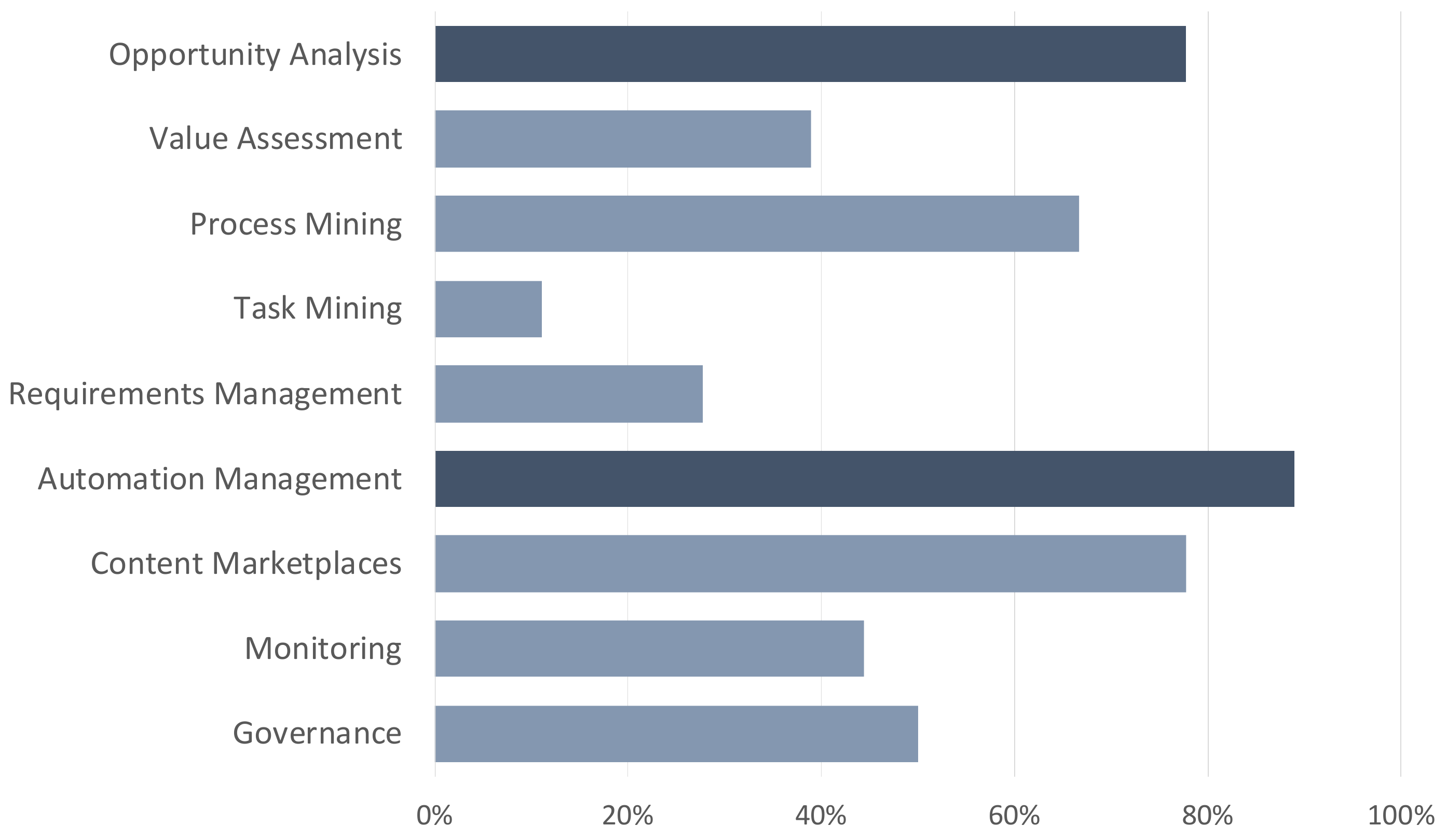}
    \caption{Vendor coverage of automation support and associated capabilities.}
    \label{fig:management}
\end{figure}

\autoref{fig:management} visualizes the indicated vendor coverage in terms of opportunity discovery and automation management, as well as their underlying capabilities.


\subsection{Opportunity Analysis}
\label{sec:results:disocvery}

Before process automation can commence, an organization must carefully consider which process(es) and which parts thereof they want to automate. Once an appropriate opportunity for automation has been identified, the details of the process and the requirements of the automation project need to be gathered.

These key activities preceding process automation can be characterized through the various associated capabilities that vendors offer:  value assessment, process mining, task mining, and requirements management. Each of these capabilities focuses on the preparation phase of an automation project and supports the discovery of automation opportunities from different perspectives. Value assessment can be seen as a business evaluation that examines whether and how much an automation benefits a company and its employees. Mining techniques can be used to gain insight into the current process execution based on the analysis of historical data. Process mining builds on the analysis of event logs to provide an understanding of the overall process. Task mining is complementary to process mining, and analyzes low-level events related to user interactions that occur during the execution of specific process tasks. Finally, requirements management focuses on detailing out all required activities associated with an automation opportunity once it has been accepted as valuable to the organization.

\mypar{Value Assessment}
The initial step of an automation project is to identify an automation opportunity that benefits the employees and the company overall. As constructing such an automation project and implementing the changes within an organization can be labor-intensive and costly, companies need to assess the potential benefits in comparison to the cost of the project. In this regard, a return on investment calculation can be performed to validate how much time a considered automation solution can actually free up for the employees and what the estimated implementation efforts are. Our analysis shows that around 40 percent of all vendors offer their customers means for value assessment. It should be noted that this preparation phase of an automation project, including the communication between business and IT, typically takes place outside automation tools. This means that oftentimes more integration efforts are required once the actual automation project starts.

\mypar{Process Mining} 
To gain a deep understanding of complex business processes and their automation potential, companies can use \textit{process mining}. By analyzing data that has been generated during the process execution (so-called event logs), companies can analyze how their processes are actually executed and, among others, identify bottlenecks, inefficiencies, and compliance issues. These insights may provide valuable input to derive and prioritize automation opportunities. We found that two-thirds of the vendors offer a process mining capability that is linked to automation. This shows that vendors focus heavily on assessing processes based on historical data to give the customer an indication for which processes an automation effort should be considered.

\mypar{Task Mining} 
Not all tasks within a process are visible in an event log. As examples, consider a user opening an email attachment or modifying an Excel spreadsheet. However, to gain a comprehensive view of a process,  these activities must also be considered. \textit{Task mining} allows one to analyze the tasks employees perform on a daily basis and helps to identify automation potential. Instead of building on event logs, task mining uses so-called user interaction logs that capture the specific mouse and keyboard interactions users engage in. Based on our analysis, only around 10 percent of vendors offer a capability to perform task mining. Compared to the offered automation capabilities of task recording (around 55 percent) and converter (around 20 percent), it seems that the focus of vendors lies more in the implementation of user task automation than the exploration and improvement of the design of user task automation.

\mypar{Requirements Management} 
Once it is decided that the automation of a specific process makes sense, the company has to start detailing the requirements for the automation project. This includes identification of the key stakeholders of the project, such as IT administrators, citizen/pro developers, and key process participants. Additionally, the requirements have to be mapped to the different automation types to identify the most suitable manner to address each requirement. This includes prioritizing, documenting, and agreeing on automation requirements with project members. 
Based on our analysis, around 30 percent of the vendors explicitly offer means for users to  manage the requirements of automation projects. That also means that around 70 percent do not offer an option to either map their requirements to an automation type nor collaborate with key stakeholders in this regard.

\subsection{Automation Management}
\label{sec:results:management}

Once a process automation initiative has been deployed, it needs to be managed during run-time. There are different capabilities that relate to managing deployed automation artifacts, primarily: monitoring, governance, and content marketplaces.

\looseness=-1 \mypar{Monitoring} 
Observing the usage and performance of automation artifacts during run-time is an important capability for most enterprise software solutions. It ensures that the automation artifact does what it is supposed to do and allows users to analyze errors of failed automation artifacts in detail. In the context of RPA, \textit{monitoring} reveals how often and when an unattended bot has run and on which server during run-time. In the context of workflow automation, it can indicate which decisions have been automated by the incorporated business rules (cf., decision automation in \autoref{sec:types:workflowautomation}) or indicate any cases that got stuck in the process.
As workflow tools work with real-time business data, having visibility into a process through  such analytical monitoring also enhances the overall understanding of process performance during run-time. Additionally, a standard capability in this category also relates to the actual management of the artifact itself, for example allowing the IT administrators to restart an automation artifact or check whether infrastructure is healthy, for example, by validating the connection to the RPA bot on a remote server. Based on the market analysis around 45 percent of vendors incorporate this capabilities in their automation offering. 

\mypar{Governance} 
Similar to the monitoring capabilities, \textit{governance} is an important aspect to manage automation artifacts. Especially organizations with many citizen developers (i.e., people having the ability to build automated solutions themselves) require a mechanism to manage and control the automation development and usage overall. This may concern managing the user authorization of employees who are allowed to build automated solutions, activate automated solutions, and participate in an automated solution during run-time by customizing company policies and rules. Furthermore, as an unattended RPA bot acts by itself once set up, there needs to be a controlling mechanism that ensures that the bots do what they are supposed to do and checks whether the bots are compliant. Therefore, similar to the software development process, the automation development needs to follow a framework that ensures its quality, re-usability, and security overall. Our analysis shows that 50 percent of vendors offer means to set up a governance structure of the automation projects within their company.

\mypar{Content Marketplaces} 
\looseness=-1 Building an automated solution can be a complex and, depending on the use case, may require a deep understanding of automation technology itself. Hence, having a content library or templates of automation artifacts of best practice use cases can significantly reduce the automation development time. Multiple automation vendors offer a marketplace or store where users can simply download and deploy pre-packaged automation solutions. This content is often managed by the vendors themselves allowing the users to solely focus on the selection of the best-fitting solution. If customization is needed, they can use automation templates and adapt it to their own requirements. Additionally, some vendors offer a private content library where automation solutions that have been built by the user can be shared privately within their organization. The vendor analysis highlights that around 80 percent of vendors focus on pre-built automation content for their customers. This highlights that many vendors want to enable their customers to quickly implement automation solutions with as least effort as possible.   

\section{Vendor Focus}
\label{sec:vendor_coverage}
In this section, we answer RQ3 by comparing the automation types, means of support, and associated capabilities that automation vendors offer. The primary purpose of our assessment is to recognize patterns in the offerings from commercial vendors and identify open opportunities. As a basis for this discussion, \autoref{table:vendorcoveragematrix} presents a vendor coverage matrix that maps the 18 vendors to the 4 automation types, 2 kinds of support (opportunity analysis and automation management), and the 21 associated capabilities. Here, we followed the vendor categorization from Gartner, which divides the vendors into so-called leaders, visionaries, challengers, and niche players according to their market position.

 \begin{table*}[!ht]
    \centering
    \footnotesize
    \setlength{\tabcolsep}{4pt}
    \begin{tabular}{rc cccc@{\hskip 2em} cccc@{\hskip 2em} cc@{\hskip 2em} ccccccccc}
    \toprule
        & & \multicolumn{4}{c}{\textbf{Leaders}} & \multicolumn{4}{c}{\textbf{Visionaries}} & \multicolumn{2}{c}{\textbf{Challeng.}} & \multicolumn{8}{c}{\textbf{Niche players}}   \\
        \midrule
         & \rotatebox[origin=l]{90}{ \textbf{Vendor coverage}} 
         & \rotatebox[origin=l]{90}{ UiPath } & \rotatebox[origin=l]{90}{ Automation Anywhere } & \rotatebox[origin=l]{90}{ Microsoft } & \rotatebox[origin=l]{90}{ Blue Prism } & \rotatebox[origin=l]{90}{ WorkFusion } & \rotatebox[origin=l]{90}{ Pegasystems } & \rotatebox[origin=l]{90}{ Appian } & \rotatebox[origin=l]{90}{ Servicetrace } & \rotatebox[origin=l]{90}{ NICE } & \rotatebox[origin=l]{90}{ EdgeVerse Systems } & \rotatebox[origin=l]{90}{ NTT } & \rotatebox[origin=l]{90}{ Samsung SDS } & \rotatebox[origin=l]{90}{ SAP } & \rotatebox[origin=l]{90}{ Nintex } & \rotatebox[origin=l]{90}{ IBM } & \rotatebox[origin=l]{90}{ Kryon } & \rotatebox[origin=l]{90}{ Cyclone Robotics } & \rotatebox[origin=l]{90}{ Laiye} \\
        \midrule
         \textbf{Capabilities per vendor} & ~ & 18	& 10 &	11 &	13 &	9 &	12 &	16 &	9 &	7 &	11 &	6	& 9 &	15 &	13 &	11 &	5 &	15 &	8 \\
        \midrule 
        \\
        \textbf{RPA} & 18 & x & x & x & x & x & x & x & x & x & x & x & x & x & x & x & x & x & x \\ 
        \midrule
        Attended RPA & 18 & x & x & x & x & x & x & x & x & x & x & x & x & x & x & x & x & x & x \\ 
        Unattended RPA & 18 & x & x & x & x & x & x & x & x & x & x & x & x & x & x & x & x & x & x \\ 
        Task Conversion & 4 & ~ & ~ & ~ & ~ & ~ & ~ & ~ & ~ & x & x & ~ & ~ & ~ & ~ & ~ & x & x & ~ \\ 
        Task Recording & 10 & x & x & x & x & ~ & ~ & x & x & ~ & ~ & x & x & x & ~ & ~ & ~ & x & ~ \\ 
        Document Processing & 13 & x & x & ~ & x & x & ~ & x & x & ~ & x & x & ~ & x & x & x & ~ & x & x \\ 
        \noalign{\smallskip}
        \textbf{Workflow Automation} & 14 & x & ~ & x & x & x & x & x & x & ~ & x & ~ & x & x & x & x & ~ & x & x \\
        \midrule
        Decision Automation & 7 & ~ & ~ & ~ & x & ~ & x & x & x & ~ & ~ & ~ & ~ & x & ~ & x & ~ & x & ~ \\ 
        Approval Support & 14 & x & ~ & x & x & x & x & x & x & ~ & x & ~ & x & x & x & x & ~ & x & x \\ 
        Escalation Modeling & 5 & x & ~ & ~ & x & ~ & ~ & x & ~ & ~ & ~ & ~ & ~ & x & x & ~ & ~ & ~ & ~ \\ 
        Automated Notification & 4 & x & ~ & ~ & ~ & ~ & ~ & ~ & ~ & ~ & ~ & ~ & ~ & x & x & ~ & ~ & ~ & x \\ 
        Case Management & 8 & x & x & ~ & ~ & x & x & x & x & ~ & ~ & ~ & ~ & ~ & x & x & ~ & ~ & ~ \\ 
        \noalign{\smallskip}
        \textbf{iPaaS} & 12 & x & ~ & x & x & x & x & x & ~ & ~ & x & ~ & x & x & x & x & ~ & x & ~ \\ 
        \midrule
        Data Mapping & 12 & x & ~ & x & x & x & x & x & ~ & ~ & x & ~ & x & x & x & x & ~ & x & ~ \\ 
        API Connectors & 12 & x & ~ & x & x & x & x & x & ~ & ~ & x & ~ & x & x & x & x & ~ & x & ~ \\ 
        \noalign{\smallskip}
        \textbf{Integrated Automation} & 9 & x & ~ & x & ~ & ~ & x & x & ~ & ~ & ~ & ~ & x & x & x & x & ~ & x & ~ \\ 
        \midrule
        Low-Code Development & 9 & x & ~ & x & ~ & ~ & x & x & ~ & ~ & ~ & ~ & x & x & x & x & ~ & x & ~ \\ 
        Mobile Integration & 7 & ~ & ~ & x & ~ & ~ & x & x & ~ & ~ & ~ & ~ & ~ & x & x & x & ~ & x & ~ \\ 
        \noalign{\smallskip}
        \textbf{Opportunity Analysis} & 14 & x & x & x & x & ~ & x & x & x & x & x & ~ & ~ & x & ~ & x & x & x & x \\
        \midrule
        Value Assessment & 7 & x & x & ~ & x & ~ & x & x & ~ & x & x & ~ & ~ & ~ & ~ & ~ & ~ & ~ & ~ \\ 
        Process Mining & 12 & x & x & x & x & ~ & x & ~ & x & x & x & ~ & ~ & x & ~ & x & x & x & ~ \\ 
        Task Mining & 2 & x & ~ & ~ & ~ & ~ & ~ & ~ & ~ & ~ & ~ & ~ & ~ & ~ & ~ & ~ & ~ & x & ~ \\ 
        Requirements Management & 5 & x & ~ & ~ & ~ & ~ & ~ & ~ & ~ & ~ & x & ~ & ~ & ~ & ~ & ~ & x & x & x \\ 
        \noalign{\smallskip}
        \textbf{Automation Management}  & 16 & x & x & x & x & x & x & x & x & x & x & x & x & x & x & ~ & ~ & x & x \\ 
        \midrule
        Monitoring & 8 & x & x & x & ~ & x & ~ & x & ~ & ~ & ~ & x & ~ & x & x & ~ & ~ & ~ & ~ \\ 
        Governance & 9 & x & x & ~ & x & ~ & ~ & x & x & x & ~ & x & x & ~ & ~ & ~ & ~ & ~ & x \\
        Content Marketplaces & 14 & x & x & x & x & x & x & x & ~ & x & x & ~ & x & x & x & ~ & ~ & x & x \\ 
        \bottomrule
    \end{tabular}
    \caption{Vendor coverage matrix of automation types, support means, and underlying capabilities.}
\label{table:vendorcoveragematrix}
\end{table*}

\mypar{Capabilities versus market position} 
When considering the overall coverage of the 21 capabilities by the different vendors, we observe clear differences between certain players that offer most capabilities (e.g., UiPath with 18, Appian with 16, and SAP and Cyclone Robotics with 15 capabilities), and players that focus on a clearly smaller set of capabilities (such as Kryon with 5 and NTT with 6 capabilities). It is interesting to observe that, even though UiPath, one of the market leaders, has the broadest offering, there is no clear relation between the vendor categories in terms of their position in the market and the (number of) capabilities they offer. For example, certain leaders offer fewer capabilities than certain visionaries, challengers, or niche players.


\mypar{Offered automation types} Looking at the coverage of automation types, we observe that half (9 out of 18) of the vendors cover all four primary automation types, whereas only 4 vendors just target RPA solutions. This confirms our initial premise that process automation goes considerably beyond RPA, even for RPA vendors.
An interesting observation here is that there is a clear pattern in terms of the automation types that vendors offer. All 14 vendors that cover 2 automation types provide workflow automation solutions next to RPA, whereas the 3 vendors that cover 3 automation types all focus on RPA, workflow automation, and iPaaS. This means that, for example, there is no vendor that just offers RPA and iPaaS, or that offers integrated process automation solutions while omitting one of the three other types from its offering. 


\mypar{Implementation over pre and post-hoc support}
Finally, we observe that 14 out of 18 vendors provide some means for the discovery of automation opportunities, most commonly by offering process mining capabilities (12 vendors). However, more vendors provide support in terms of automation management capabilities (16 vendors). This is most commonly achieved through content marketplaces, which allow organizations to re-use common automation patterns.  The provision of post-implementation support through monitoring and governance capabilities is less common, with just 8 and 9 vendors, respectively.
This analysis reveals that vendors are primarily focused on the actual implementation phase of automation projects and not yet as much on the early phases, i.e., the discovery of automation opportunities, or the post-hoc support.

\section{Conclusion}
\label{sec:conclusion}
Finally, we conclude our structured market analysis on business process automation vendors by providing key insights related to our research questions and include an outlook and discussion for future research.

\mypar{Automation is more than RPA} Even though RPA is a key component of all automation vendors, our analysis for RQ1 shows that it is not the only automation type these vendors focus on. As automation requirements can vary significantly depending on the use case, automation vendors also need to focus on additional capabilities to meet them. A workflow, where a human interacts with an automation artifact to perform a user task, an API integration to connect to a system, or the ability to build an application where a user can provide data to a process, are just some examples of automation that are beyond the capabilities of RPA. 
Therefore, in this context, it is important to understand which automation capability or combination thereof should be used for a given automation scenario. 
Such as, for instance, combining the ability of RPA to automate individual process steps with workflow automation to coordinate the overarching process and iPaaS to ensure that all relevant systems are updated accordingly.


\mypar{Automation is intended for everyone} 
\looseness=-1 Insights obtained for RQ2 show that most vendors want their automation capabilities to be used by anyone within a company, regardless of their technical expertise. Therefore, vendors put a strong emphasis on user experience and re-usable building blocks to move away from pro-code development environments that traditionally require users to have a deep understanding of technology. Instead, the focus shifts towards the business user and process owners being the key stakeholder within the overall automation project. This has several implications. 

First, business users and process owners have deep insights into their processes and can therefore assess quite well for which processes automation is viable. Second, the use of no-code/low-code allows business users and process owners to develop automation artifacts themselves, which breaks up the traditional dependency constellation between business and IT, where IT builds a solution that is then consumed by the business side. Third, this situation creates an increasing need for automation governance. When many employees can create automation artifacts, a framework has to be setup that clearly manages who is allowed to build those artifacts and for which processes. Among others, business users by themselves will not always be able to anticipate the technical implications of their automation artifact or how it relates to other, already existing solutions.

\mypar{Vendors have different profiles} RQ3 revealed that each vendor has a different focus when it comes to their strategic positioning. While some vendors have a stronger focus on the opportunity analysis and provide respective capabilities, others provide more capabilities in the area of workflow automation and iPaaS. 
Additionally, some vendors also bring intelligence to their automation portfolio by combining their automation capabilities with artificial intelligence techniques. As examples, consider performing sentiment analysis on incoming emails or automatically classifying customer service requests. As discussed earlier, Gartner divides the vendors into leaders, visionaries, challengers, and niche players according to their market position. However, our analysis reveals that the vendors' market position cannot be explained by the number of capabilities they offer, with some niche players offering considerably more capabilities than certain leaders. This implies that the quality of a vendor's offering is more important than its breadth in terms of quantity.

\mypar{Trends and outlook} Our analysis has revealed two major trends. First, we observe that many vendors aim to offer capabilities that go beyond RPA and focus on integrating multiple automation capabilities in a single automation suite. Second, we can see an increasing interest in integrating machine learning and other techniques from artificial intelligence into existing automation capabilities. In this way, also less structured and potentially even cognitive tasks can be automated. Vendors typically use terms such as \textit{intelligent automation} or \textit{hyperautomation} to refer to this combination.    

Our work also provides explicit input for both industry and research with respect to future directions. As for \textit{industry}, our analysis shows that there is more need for automation support in both the pre and post-automation phases. Due to the increasing adoption of automation solutions, increased demand monitoring and governance capabilities can be expected. Our analysis has also revealed that the quality of the offered capabilities seems to be more relevant for the overall market position. As for \textit{research}, we would like to highlight three main points. First, there is a need to understand to what extent the offered capabilities are actually adopted by users and which challenges come with that. While such analyses have been conducted for RPA specifically~\cite{van2018robotic,hofmann2020robotic,enriquez2020robotic}, also other facets of automation need to be considered. Second, the increasing introduction of automation artifacts raises the question of how humans and these artifacts actually work together. While this is generally not a new topic~\cite{janssen2019history}, there is a need to better understand this in the context of business process automation. Third, there is potential to extend the scope of RPM techniques ~\cite{leno2020identifying,agostinelli2021exploring}. It would be interesting to explore whether similar technical support can be also provided for other types of automation, such as iPaaS.


\bibliographystyle{apalike}

\end{document}